\documentclass{article}
\usepackage{arxiv}

\usepackage[utf8]{inputenc} 
\usepackage[T1]{fontenc}    
\usepackage{hyperref}       
\usepackage{url}            
\usepackage{booktabs}       
\usepackage{amsfonts}       
\usepackage{nicefrac}       
\usepackage{microtype}      
\usepackage{lipsum}

\usepackage{appendix}
\usepackage{amsmath}
\usepackage{graphicx}
\usepackage{lineno}
\usepackage{array}
\usepackage{longtable}
\usepackage{algorithm}
\usepackage{algpseudocode}
\usepackage{amssymb}
\usepackage[absolute,overlay]{textpos}

\usepackage{enumitem}
\usepackage{url}
\usepackage{adjustbox}
\usepackage[absolute,overlay]{textpos}
\usepackage{graphics}

\usepackage{mathrsfs} 

\usepackage{stmaryrd}
\usepackage{amsthm}
\usepackage{amssymb}

\title{Low-dimensional offshore wave input for extreme event quantification}

\author{
  Kenan \v Sehi\'c\thanks{Corresponding author: kense@dtu.dk} \\
  Department of Applied Mathematics and Computer Science\\
  Technical University of Denmark\\
  DK-2800 Kgs. Lyngby, Denmark\\
   \And
 Henrik Bredmose\\
  Department of Wind Energy\\
  Technical University of Denmark\\
  DK-2800 Kgs. Lyngby, Denmark\\
  \And
 John D. S\o rensen\\
  Department of Civil Engineering \& Department of Wind Energy\\
  Aalborg University \& Technical University of Denmark\\
  Denmark\\
  \And
 Mirza Karamehmedovi\'c\\
  Department of Applied Mathematics and Computer Science\\
  Technical University of Denmark\\
  DK-2800 Kgs. Lyngby, Denmark\\
}

\begin{document}
\maketitle

\begin{abstract}
In offshore engineering design, nonlinear wave models are often used to propagate stochastic waves from an input boundary to the location of an offshore structure. Each wave realization is typically characterized by a high-dimensional input time series, and a reliable determination of the extreme events is associated with substantial computational effort. As the sea depth decreases, extreme events become more difficult to evaluate. We here construct a low-dimensional characterization of the candidate input time series to circumvent the search for extreme wave events in a high-dimensional input probability space. Each wave input is represented by a unique low-dimensional set of parameters for which standard surrogate approximations, such as Gaussian processes, can estimate the short-term exceedance probability efficiently and accurately. We demonstrate the advantages of the new approach with a simple shallow-water wave model based on the Korteweg-de Vries equation for which we can provide an accurate reference solution based on the simple Monte Carlo method. We furthermore apply the method to a fully nonlinear wave model for wave propagation over a sloping seabed. The results demonstrate that the Gaussian process can learn accurately the tail of the heavy-tailed distribution of the maximum wave crest elevation based on only $1.7\%$ of the required Monte Carlo evaluations.
\end{abstract}

\keywords{sequential design \and Gaussian process regression \and offshore applications \and dimensionality reduction \and extreme events}
\newpage
\section{Introduction}
\label{intro}
The occurrence of extreme events is typically quantified by a $d$-fold integral, called the probability of failure,
\begin{equation}\label{pf}
    P_F = \int_{\gamma-g(\theta)\leq0} \pi_d(\theta) d\theta,
\end{equation}
where $\theta \in \mathbb{R}^{d}$ is the stochastic input of a limit-state function $\gamma-g(\theta)$, $\pi_d$ is the joint probability density function (PDF) for $\theta$ and $g(\theta)\geq \gamma$ defines a failure event with a failure threshold $\gamma$. In the present study, the failure event is defined by a wave crest elevation $g(\theta)$ (for example, the maximum 1-hour crest elevation) exceeding a certain threshold $\gamma$. We recognize the probability of failure, Eq.~\eqref{pf}, as the short-term exceedance probability since the failure event is defined for a predefined sea state. For a real design situation, the short-term probability distribution would be further integrated over the range of possible sea states to define the long-term exceedance probability. We here focus only on the efficient evaluation of the short-term exceedance probability. We define $\theta$ to be a highly dimensional vector of stochastic Fourier coefficients that generate input waves which recreate real offshore conditions \cite{bigoni,samocita,samocita2,turk}.

In offshore engineering, the short-term exceedance probability, Eq.~\eqref{pf}, is often estimated with the well-known reliability approaches FORM/SORM, which are the first/second-order Taylor series approximations of the limit-state function at the design point. However, FORM/SORM idealize the failure surface and fall short of defining multiple failure regions \cite{form}. Additionally, as a geometrical approach, FORM/SORM do not provide error estimates. A robust alternative is the simple Monte Carlo (MC) method \cite{mcbook}. It does not depend on the dimension of the input and can find multiple design points for almost any numerical model. In this framework, the short-term exceedance probability, Eq.~\eqref{pf}, is defined as the sample mean of the indicator function $\mathbb{I}(\theta)$, where $\mathbb{I}(\theta) = 1$ if $g(\theta)\geq \gamma$ and $\mathbb{I}(\theta) = 0$ otherwise. The idea is to compute $N$ blind samples of the numerical model and estimate the sample mean. While being a straightforward approach to implement, MC is infeasible in conjunction with expensive numerical models due to the slow convergence rate $\mathcal{O}(N^{-1/2})$. As we are only able to approximate the sample mean Eq.~\eqref{pf}, we employ the mean squared error measure to define a sufficient number of evaluations $N$ for a prescribed relative error \cite{mcbook}. For example, the response value of an exceedance probability of $2 \cdot 10^{-3}$ with the relative error less than $0.1$ requires at least $N=5\cdot 10^4$ evaluations. A numerical model which needs 1 minute to compute would require approximately 35 days to find the short-term exceedance probability. Therefore, surrogate methods such as Polynomial Chaos expansion \cite{xiuli} and Gaussian (Kriging) process \cite{bruno} are usually proposed to define a low-cost surrogate for the numerical model using a small number of evaluations. The low-cost surrogate is then employed to approximate efficiently the short-term exceedance probability. However, the requirements for standard surrogate methods typically exponentially increase with the dimension due to the curse of dimensionality~\cite{paul1}. For high-dimensional problems, as the one considered in this paper, this approach becomes infeasible.

Alternative approaches consider only the statistical characteristics of extreme events such as extreme value theorems \cite{evt} and large deviations theory \cite{ld2,ld}. However, these methods have essential limitations requiring various extrapolation schemes due to the insufficient size of the sample set, and they therefore cannot always explain the non-trivial shape of the tail. The Fokker-Planck equation (FPE) \cite{fpe} incorporates into the estimation the governing dynamical system of the model, as well as the stochastic nature of extreme events. However, even in low-dimensional cases, the FPE is prohibitively difficult to solve.

Mohamad et al. \cite{mo} proposed a sequential sampling strategy based on Gaussian process regression and statistical properties of extreme events. They formulated an optimization process that selected the most informative design point according to the log-L2 criterion, which eventually reduce the uncertainties in the prediction of extreme events. To avoid the curse of dimensionality, their numerical implementation is low-dimensional. For a wave-propagation application, they characterize the input condition by just two parameters, namely wave group length and amplitude. Although the wave group parameters are good indicators of extreme wave events \cite{sapsis,asa,wavepack2,wavepack1,wavepack3}, the propagation over long distances from the generation boundary makes single-group characteristics less relevant in the general case. Other options would be to employ standard dimensionality reduction tools such as principal component analysis (PCA) \cite{pcabook}, partial least squares regression (PLS) \cite{pls1,pls4}, active-subspace analysis \cite{paul1,asa}, or autoencoders \cite{vae_var}. However, these standard dimensionality reduction tools may provide inefficient reduction in offshore applications or suffer from expensive and intrusive implementation.

To improve on this, we propose a low-dimensional classification representation of the high-dimensional Gaussian input. Based on the input sample set of the predefined input sea state, we classify the wave input by classification parameters unique for each time series. The classification parameters are now the low-dimensional design input parameters for Gaussian process regression. To reduce the prediction uncertainties for extreme events, we select the most informative design points from the input Gaussian evaluations based on the probability of misclassification \cite{bruno}. We test our approach in two numerical experiments. The first is based on the Korteweg-de Vries equation, which is a simple shallow-water wave model for which we can generate the reference solution with the simple Monte Carlo method. The second case is fully nonlinear wave propagation over a sloping seabed, computed with the OceanWave3D model \cite{allan}. We select the classification parameters heuristically, and view this paper as initial work on the low-dimensional representation of offshore wave input, or generally of input time-series for any system.

Section 2 covers Gaussian process regression in general. We introduce the low-dimensional representation of input time series in Section 3, and present the numerical experiments in Section 4. There, we demonstrate that the low-dimensional representation based on our classification parameters is suitable for the Gaussian process design and efficient quantification of extreme events. We present our conclusions in Section 5.

\section{Gaussian process regression}
When the numerical cost of a model prohibits the quantification of the involved uncertainties, it is natural to try to construct a surrogate model, a low-cost replacement based on a small number of evaluations $N_{\rm GP}$ of the original model. In this work we use Gaussian or Kriging process regression (GP), which employs the Gaussian distribution over a training set based on a Bayesian approximation \cite{rasmussen}. The training set, $\mathcal{S} = \{(\theta_i, g(\theta_i))\}$, includes the matrix of the input parameters $\mathbf{X} = (\theta_i) \in \mathbb{R}^{N_{\rm GP} \times d}$ and the corresponding evaluations of a numerical model $Y = (Y_i = g(\theta_i)) \in \mathbb{R}^{N_{\rm GP}\times 1}$. For an arbitrary smooth function $g$, we define a Gaussian process approximation by \cite{rasmussen,bruno}

\begin{equation}
    g(\theta) \approx \hat{g}(\theta) = \beta^T \cdot f_{\rm T}(\theta) + \sigma^2 Z(\theta,\omega_z),
\end{equation}
where $\beta^T \cdot f_{\rm T}(\theta)$ is the trend, $\sigma^2$ is the variance of the Gaussian process, and $Z(\theta,\omega_z)$ is a zero-mean, unit-variance stationary Gaussian process with $\omega_z \in \Omega$ as an elementary event in the probability space $(\Omega, \mathcal{F}, \mathbb{P})$. The trend $f_T(\theta)$ describes the global behavior of the training set, and is defined using simple regression. The complexity of the Gaussian process $Z$ is described by a stationary kernel matrix $\mathbf{K}_{ij} = K(|\theta_i - \theta_j|;\Theta)$, where $\Theta$ are the hyperparameters such as the overall correlation of the samples or the smoothness of the training set. The performance of the GP regression depends highly on the selection of the kernel function. For example, a periodic kernel may be best suited for a periodic training set. We define the regression parameters $\beta$ and $\sigma^2$ based on generalized least-squares regression \cite{bruno}. The hyperparameters $\Theta$ for the kernel matrix $\mathbf{K}_{ij}$ are found using maximum likelihood estimation. A typical challenge is to select an optimal size $N_{\rm GP}$ of the training set. Gramacy and Apley \cite{gramacy} propose to select the training size by adding observations until a certain mean squared predictive error is no longer fulfilled. Once the GP model is trained, we estimate for an arbitrary input sample $\theta^*$ the prediction $\mu_g (\theta^*)$ and the variance (i.e., an uncertainty measure) $\sigma_g^2(\theta^*)$ via \cite{rasmussen,bruno}

\begin{equation}\label{mugp}
    \mu_g(\theta^*) = f_{\rm T}(\theta^*)\cdot\beta + k(\theta^*)^T\mathbf{K}^{-1}(Y - \mathbf{F}_{\rm T}\beta),
\end{equation}
and
\begin{equation}\label{sdgp}
    \sigma_g^2(\theta^*) = \sigma^2\Bigg( 1 - \Big\langle f_{\rm T}(\theta^*)^Tk(\theta^*)^T \Big\rangle \begin{bmatrix}
0 & \mathbf{F}_{\rm T}^T\\
\mathbf{F}_{\rm T} & \mathbf{K} \end{bmatrix}^{-1} \begin{bmatrix}
f_{\rm T}(\theta^*) \\
k(\theta^*) \end{bmatrix} \Bigg).
\end{equation}
Here $k(\theta^*)$ is the correlation between the arbitrary input sample and the rest of the samples within the training set, and $\mathbf{F}_{\rm T}$ is the information matrix regarding the GP trend. 
\subsection{Sequential Uncertainty Reduction for Extreme Events}\label{sur}

Gaussian process regression provides the posterior distribution, based on the moments~\eqref{mugp} and~\eqref{sdgp}, which describes the quality of the predictions and, in general, the performance of the surrogate model. An arbitrary Gaussian prediction $\mu_g(\theta^*)$ and the corresponding standard deviation $\sigma_{g}(\theta^*)$ define the confidence interval (CI) \cite{bruno,rasmussen}
\begin{equation}\label{ci}
    \mu_{CI}^{\pm} (\theta^*) = \mu_g(\theta^*) \pm \alpha \cdot \sigma_{g}(\theta^*),
\end{equation}
with confidence level $\alpha$. For example, $\alpha=1.96$ for $95\%$ confidence, and $95\%$ of the area of the normal distribution is within 1.96 standard deviations. Narrower confidence intervals mean more confidence in the predictions and generally in the performance of the surrogate model, under the assumption that the limit-state function is smooth. However, a significantly narrow CI can still over/under-predict a highly nontrivial response, because Gaussian process regression is a design method. The training set and the initial design may not be sufficient to accurately describe a complicated function. The application is not straightforward and should be employed carefully with an understanding of the numerical model and the data itself. Hence, the idea is to sequentially design numerical experiments $\theta^*$ for training while exploring optimally the probability space (i.e., using a learning function) to create an optimal Gaussian process. Mohamad et al. \cite{mo} proposed a sequential design utilizing the log-L2 distance between the upper and lower bound of CI to reduce uncertainties for extreme events. Sch\"obi et al. \cite{bruno} used the probability of misclassification to define which numerical experiments $\theta^*$ are close to a predefined failure threshold or are poorly predicted. We find this sufficient for our analysis.

As the probability of failure essentially involves a binary classification, with 1 for failure and 0 otherwise, the misclassification of predictions based on the first two moments and on the failure threshold $\gamma$ is a natural option for Gaussian process regression. We write for the probability of misclassification $P_M$ \cite{bruno}

\begin{equation}\label{pm}
    P_M (\theta^*) = \Phi\left(-\frac{|\mu_g(\theta^*) - \gamma|}{\sigma_g(\theta^*)}\right).
\end{equation}
Here, the $U$-function (the learning function) is the fraction in the argument in~\eqref{pm}. It is recognized as the reliability index attached to $P_M$ with $\Phi$ as the cumulative distribution function of the standard normal distribution. Small values of the $U$-function reveal the samples $\theta^*$ that are close to the predefined failure threshold $\gamma$ or have high uncertainties in predictions. Therefore, to improve the training set, we select the most informative design points, i.e., the input parameters $\theta^*$ with the smallest $U$-values, based on the generated samples $\mathcal{S}_{\rm MC} = \{\theta_1,\cdots, \theta_N\} \in \mathbb{R}^{N\times d}$. This produces a Gaussian process designed specially for extreme events. The procedure is iterative, as we add more $U$-based design points $\theta^*$ until an error measure based on the confidence interval for the failure level drops below some prescribed threshold.

\begin{algorithm}
\caption{Gaussian discrete learning with the $U$-function}
\label{seq}
\begin{algorithmic}[1]
\Procedure{$U$-GP}{$\theta \in \mathbb{R}^d$- the input parameters, $g$ - the numerical model, $\mathcal{S} = \{(\theta_i, g(\theta_i))\} \in \mathbb{R}^{N_{\rm GP}\times d}$ - the training set, $\mathcal{S}_{\rm MC} = \{\theta_1,\cdots, \theta_N\} \in \mathbb{R}^{N\times d}$ - the generated samples, $\gamma$ - the failure threshold}
    \State Train a Gaussian process on the training set $\mathcal{S}$.
    \State Use the trained GP to estimate the first two moments for the generated samples $\mathcal{S}_{\rm MC}$ following Eq.~\eqref{mugp} and Eq.~\eqref{sdgp}.
    \State Estimate the confidence interval based on $\mathcal{S}_{\rm MC}$ with Eq.~\eqref{ci} and quantify the exceedance probabilities $\hat{P}_F$ and $\hat{P}_F^{\pm}$ as the sample means of the indicator function based on the failure threshold $\gamma$.
    \State Evaluate the error measure $\varepsilon_{\gamma}$ by Eq.~\eqref{er} for the failure threshold $\gamma$.
    \Repeat
    \State For each generated sample $\theta^*$ from the set $\mathcal{S}_{\rm MC}$, estimate the $U$-value based on the $U$-function and the failure threshold $\gamma$,
    \[
    U(\theta^*) = \frac{|\mu_g(\theta^*) - \gamma|}{\sigma_g(\theta^*)}.
    \]
    \State Select an arbitrary number of samples $\theta^*\in \mathbb{R}^{N_U\times d}$ that achieve minimal $U$-values and evaluate a numerical model $g(\theta^*)$.
    \State Include the pair $(\theta^*,g(\theta^*))$ within the training set $\mathcal{S}$.
    \State Repeat Lines 2-5.
    \Until{For a prescribed error threshold $\xi$: $\varepsilon_{\gamma} \leq \xi$.}
\EndProcedure
\end{algorithmic}
\end{algorithm}
The procedure is described in more detail in \textbf{Algorithm \ref{seq}}. It can be seen as discrete learning, since we rely on the generated samples $\mathcal{S}_{\rm MC}$. The optimal approach is to select the sample that maximizes the probability of misclassification or minimizes the $U$-function based on the trained GP. As the trained GP is a surrogate model, the optimization procedure is low-cost. However, for our approach, which uses classification parameters, we can only build the procedure on the generated samples. The error measure, a stopping criterion, is defined as \cite{bruno}
\begin{equation}\label{er}
    \varepsilon_{\gamma} = \frac{|\hat{P}_F^+ - \hat{P}_F^-|}{\hat{P}_F},
\end{equation}
where the exceedance probabilities $\hat{P}_F$ and $\hat{P}_F^{\pm}$ are estimated as the sample means of the indicator function by the simple Monte Carlo method based on the generated samples $\mathcal{S}_{\rm MC}$. The exceedance probabilities $\hat{P}_F^{\pm}$ involve the upper and lower bounds of the confidence interval. The error measure~\eqref{er} usually ranges between $0.5$ and $2$ \cite{bruno}. It depends on the final failure level and the accuracy requirement. If the error measure is not achieved (i.e., $\varepsilon_{\gamma} > \xi$), the discrete learning converges to the simple Monte Carlo estimation of the generated samples. 

\section{Low-dimensional representation}
When the input dimension $d$ exceeds extreme values, e.g., $d=100$, it is required to define a larger training set $N_{\rm GP}$ based on the factorial design. However, the computation of the Gaussian process becomes impractical as an $N_{\rm GP} \times N_{\rm GP}$ kernel matrix needs to be inverted several times to make an arbitrary prediction, which costs $\mathcal{O}(N_{\rm GP}^3)$. Therefore, the literature on Gaussian process regression typically covers low-dimensional numerical implementations~\cite{sapsis,mo}.

We define the input parameters within a high-dimensional probability space to recreate the environmental conditions accurately. The boundary condition, which we note as the wave input at $x=0$, is defined as a Fourier series with multi-dimensional random coefficients drawn from the standard normal distribution \cite{naess}

\begin{equation}\label{bc}
    \eta(x,t)=\sum^{d/2}_{j=1}\sqrt{S(f_j)\cdot \Delta f}\Bigg[A_{j}\cos(2\pi f_jt-k_jx)+B_j\sin(2\pi f_jt-k_{j}x)\Bigg].
\end{equation}
Here $S(f_j)$ is the wave energy spectrum, i.e., JONSWAP \cite{turk}, $f_j$ are wave frequencies, $k_j$ are the wavenumbers and $A_j$ and $B_j$ are random variables drawn from the standard normal distribution $\mathcal{N}(0,1)$. In this paper, the input parameters $A_j$ and $B_j$ are written $\theta=(A_1,\dots,A_{d/2},B_1,\dots,B_{d/2}) \in\mathbb{R}^d$. The dimensionality $d$ is based on the frequency step $\Delta f = 1/T$, where $T$ is the time duration of the numerical simulation. For example, for 1-hour wave propagation we have $d = 1802$. This is an extreme problem of uncertainty quantification due to the curse of dimensionality, which typically makes the problem complexity grow exponentially with the dimension. Also, in offshore applications the numerical model usually imposes intensive computation and is to be employed only a limited number of times.~\cite{mo}

We propose a novel approach to quantify extreme events in otherwise infeasible situations. The low-dimensional representation approach classifies initial surface elevations based on non-dimensional classification parameters as each surface elevation (i.e. time-series) has a unique set of characteristics. We change the problem set from the extreme high-dimensional Fourier coefficients $\theta$ to the low-dimensional representation based on the classification parameters $K$. In the present study, we employ standard statistical measures \cite{low} to classify the wave input Eq.~\eqref{bc}. Following the simple Monte Carlo requirement for the exceedance probability of $2\cdot 10^{-3}$, we first generate $N=5\cdot10^{4}$ initial surface elevations for the time duration $T$, and classify the wave input heuristically based on $10$ classification parameters $K_{1-10}$ (scaled by their maximum values) as \cite{low}
\begin{itemize}[label={•}]
    \item $K_1$ - the maximum crest elevation at $x=0$.
    \item $K_2$ - the wave input variance - second moment $\sigma^2$.
    \item $K_3$ - the wave input skewness - third moment.
    \item $K_4$ - the wave input kurtosis - fourth moment.
    \item $K_5$ - the wave input root mean square (RMS).
    \item $K_6$ - the wave input approximate entropy, which measures complexity.
    \item $K_7$ - the wave input percentile for $50\%$.
    \item $K_8$ - the wave input percentile for $75\%$.
    \item $K_9$ - the wave input percentile for $90\%$.
    \item $K_{10}$ - the wave input mode as the most frequent value within the surface elevation.
\end{itemize}

\section{Numerical experiments}
We illustrate the proposed approach on two offshore problems. The first example uses the modified Kortweg-de Vries equation (KdV22) \cite{asa}. The second application, involving a simple OceanWave3D benchmark with wave propagation over a slope \cite{bo}, demonstrates the applicability of the proposed classification approach to a fully nonlinear model. As the computations are intensive, the quantification of extreme events using standard methods is infeasible. Generally, we assume that the computational budget is limited.

The quantity of interest is the maximum crest elevation $\eta_{\rm max}$ at the reference location $x^*$ for an offshore structure,
\begin{equation}
    \eta_{\rm max} = \max\{ \eta(x^*,t), 0\leq t \leq T \}. 
\end{equation}
The objective is to estimate the short-term exceedance probability for $\eta_{\rm max}$ based on a predefined sea state. For both cases, we employ the JONSWAP spectrum, which is typically used for extreme events analyses. Due to the computation limitations, we employ short numerical simulations to examine the advantages and the disadvantages of our approach. The calculations are executed on a personal laptop with Intel Core i5-6200U CPU @ 2.30GHz $\times$ 4.

\subsection{A simple shallow-water wave model}
Let us consider unsteady water waves defined by the Korteweg-de Vries equation (KdV) for one-dimensional nonlinear surface flows under the influence of gravity \cite{henrik,asa},
\begin{equation}\label{KdV22}
    \eta_{t}(x,t)+\sqrt{g\cdot h}\cdot\eta_{x}(x,t)+ \frac{3}{2} \sqrt{\frac{g}{h}}\eta(x,t)\eta_{x}(x,t)+\\+(\beta+\frac{1}{6})\sqrt{\frac{g}{h}}h^{3}\eta_{xxx}(x,t)+\beta h^{2}\eta_{xxt}(x,t)=0.
\end{equation}
Here $\eta(x,t)$ is the free surface elevation, measured upwards from the still water level, $\beta=19/60$, $h$ is the seabed depth, $x$ is the horizontal coordinate, $t$ is time, and $g$ is the gravitational acceleration. The flow is assumed to be inviscid and irrotational, and the seabed flat with a depth of $h=20$m. We propagate waves for $T=600$s with $\Delta t = 0.0824$s and a high-cut frequency value of $0.3$Hz. The present version of the equation with $\beta = 19/60$ provides improved dispersion properties with a Pad\'e[2,2] fit of the linear phase speed. We select $H_S=6.8$m and $T_P=15$s, as representative values of a 100-year return period.

\begin{figure}
    \centering
    \includegraphics[scale=0.2]{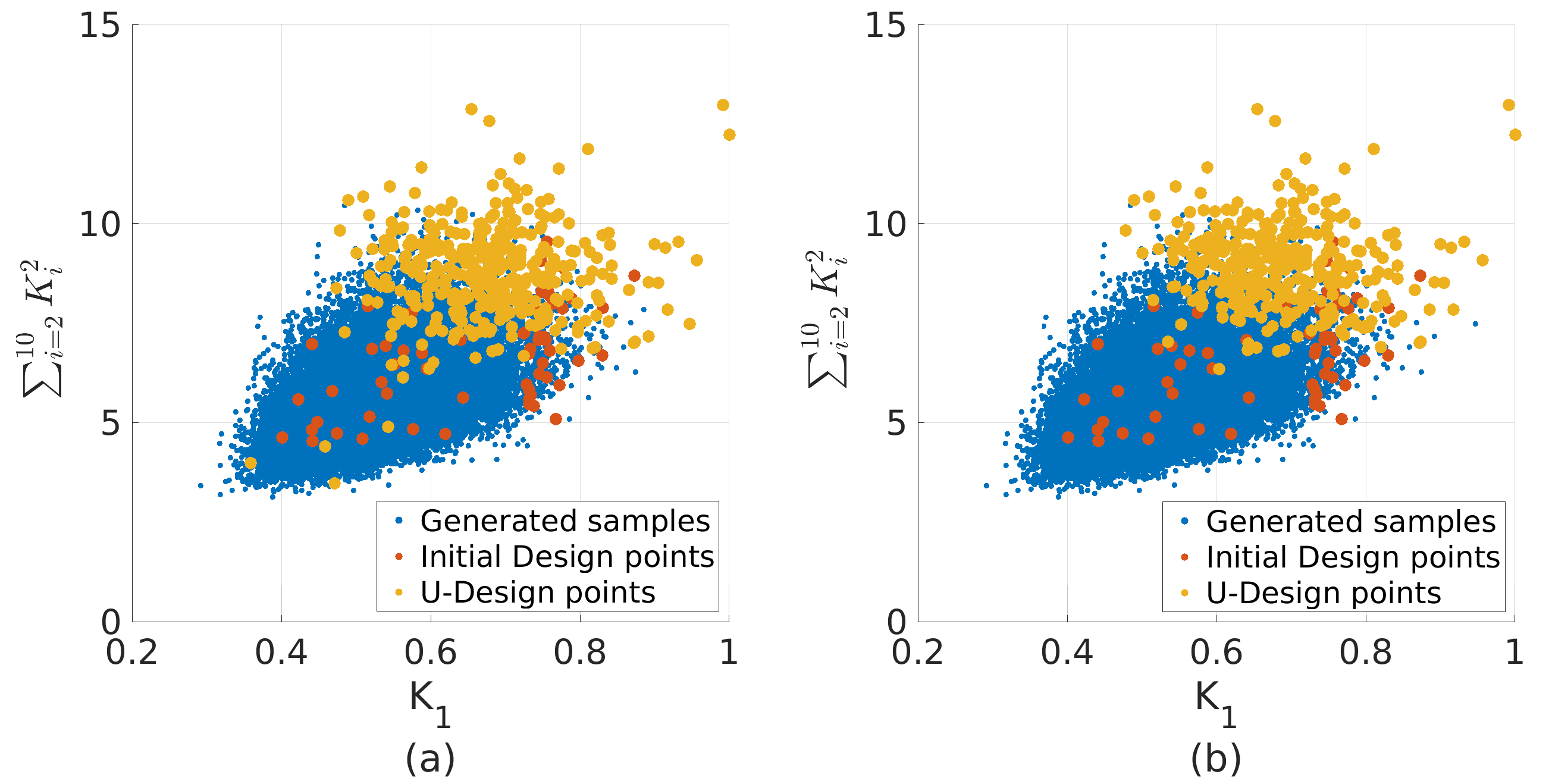}
    \caption{The wave input classification data generated using $H_S=6.8$m and $T_P=15$s. (a) $U$-design points for the pure quadratic trend, and (b) $U$-design points for the linear trend.}
    \label{kdv22:data}
\end{figure}

First, we generate $N=5\cdot 10^{4}$ wave input samples with the predefined sea state. For each wave input, we estimate the scaled classification parameters $K_{1-10}$. To visualize the data in a two-dimensional figure, we plot the sum of the squares of the classification parameters along the vertical axis, and the scaled classification parameter $K_1$ of the input maximum crest elevation $\eta_{\rm max}(0,t)$ along the horizontal axis, see Fig. \ref{kdv22:data}. Each blue sample is one wave input (i.e., time-series). On Fig. \ref{kdv22:data}, orange points (i.e., initial design points) are used for initial training of a surrogate model, while yellow points (i.e., $U$-design points) are used for active learning. Figure \ref{kdv22:data} shows the correlation between $K_1$ and the squared sum of the classification parameters. It is similar to the correlation between the wave group amplitudes and lengths in the work of Mohamad et al. \cite{mo}.

\begin{figure}
    \centering
    \includegraphics[scale=0.24]{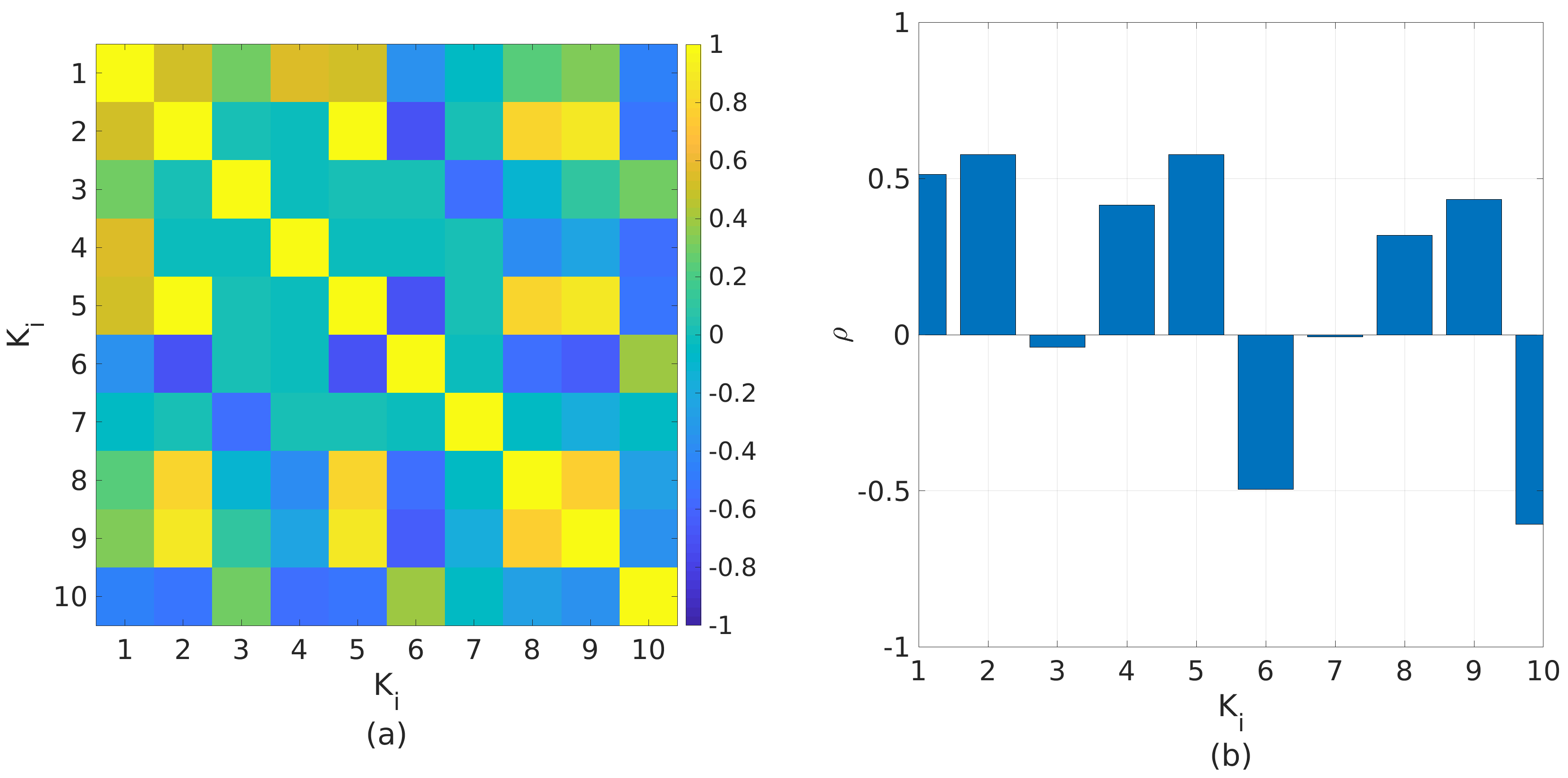}
    \caption{(a) The correlation matrix for the wave generation classification parameters $K_{1-10}$ at $x=0$. (b) The correlation relations between the quantity of interest $\eta_{\rm max}$ and the classification parameters $K_{1-10}$ at the reference location $x=x^*$.}
    \label{kdv22:corr}
\end{figure}

Figure \ref{kdv22:corr}a shows the Pearson correlations between the classification parameters. It is worth noting that the variance factor $K_2$ has a significant positive correlation with the root mean squared factor $K_5$ and the percentile parameters for $75\%$ and $90\%$ ($K_8$ and $K_9$) as expected, while it has a significant negative correlation with the approximated entropy factor $K_6$. We find the global sensitivity of the 10 classification parameters $K_{1-10}$ using the Pearson correlation and $N$ reference evaluations of the KdV22 model, see Fig. \ref{kdv22:corr}b. The term \textit{correlation} here concerns the intensity and direction of the linear relationship between two parameters. The most important parameter turns out to be the most frequent value $|K_{10}|$ within the wave input time-series. Typically, in offshore engineering, the skewness is recognized to be important for the quantification of extreme events. In this case, the relation is insignificant. We find that the least relevant parameter is the time-series percentile for $50\%$, with virtually zero Pearson correlation.

Using the factorial design, the construction of a quadratic response surface requires at least $N_{\rm GP} = (d+1)(d+2)/2$ observations \cite{Conrad:2016}. We use this relation as the reference point for the Gaussian process design. For the 10-dimensional case, the number of samples is $N_{\rm GP}=66$. However, as we focus on the sequential improvement of Gaussian models, we can select $N_{\rm GP}=60$ wave input samples randomly. As seen in Fig. \ref{kdv22:data}a, we choose to have the wave input samples equally spread.

\begin{figure}
    \centering
    \includegraphics[scale=0.3]{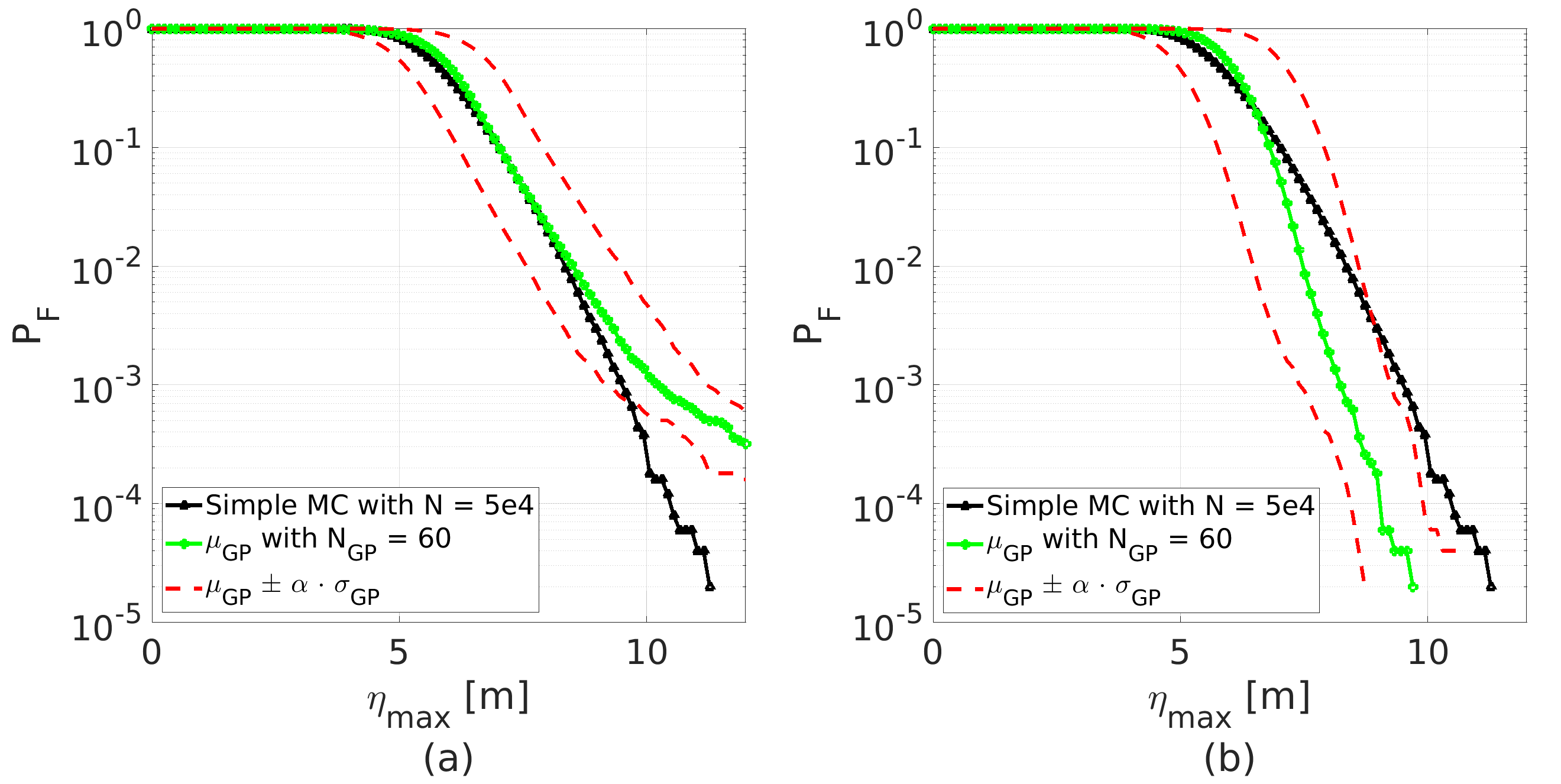}
    \caption{The initial estimation of the short-term exceedance probability $\hat{P}_F$ for the KdV22 model using the Gaussian surrogate model with (a) the quadratic trend, and (b) the linear trend.}
    \label{kdv22:pf}
\end{figure}

For each wave input, we evaluate the KdV22 model and collect the output value $\eta_{\rm max}$ to construct the training set $\mathcal{S}_K:=\{(K_{1-10, j}, \eta_{\rm{max},j}), \quad j= 1,\dots,N_{\rm GP}\}$. Using the training set $\mathcal{S}_K$, we define the Gaussian process based on the anisotropic squared exponential kernel with the pure quadratic and the linear trend, respectively. We evaluate the $N=5\cdot 10^{4}$ generated wave input samples to estimate the short-term exceedance probability, see Fig. \ref{kdv22:pf}.
There, we plot the results obtained by the initial Gaussian process with the pure quadratic and linear trend with $N_{\rm GP} = 60$. The performance of these processes are compared with the simple Monte Carlo method with $N=5\cdot10^4$ evaluations. As the KdV22 model is simple, we can generate $N=5\cdot10^4$ realizations within one day. The initial Gaussian models are inadequate for extreme events, with a pronounced discrepancy occurring for $P_F\leq 10^{-2}$. Typical offshore reliability requirements for exceedance probabilities are in the range $10^{-4}$ to $10^{-3}$. The initial value of the stopping criterion in Eq.~\eqref{er} is estimated at $\varepsilon^{\rm pure}_{9.45} = 10.96$ and $\varepsilon^{\rm lin}_{9.45} = 10.96$, respectively, for each of the trends and for $\gamma=9.45$m, see  Fig. \ref{kdv22:pf}. 

\begin{figure}
    \centering
    \includegraphics[scale=0.3]{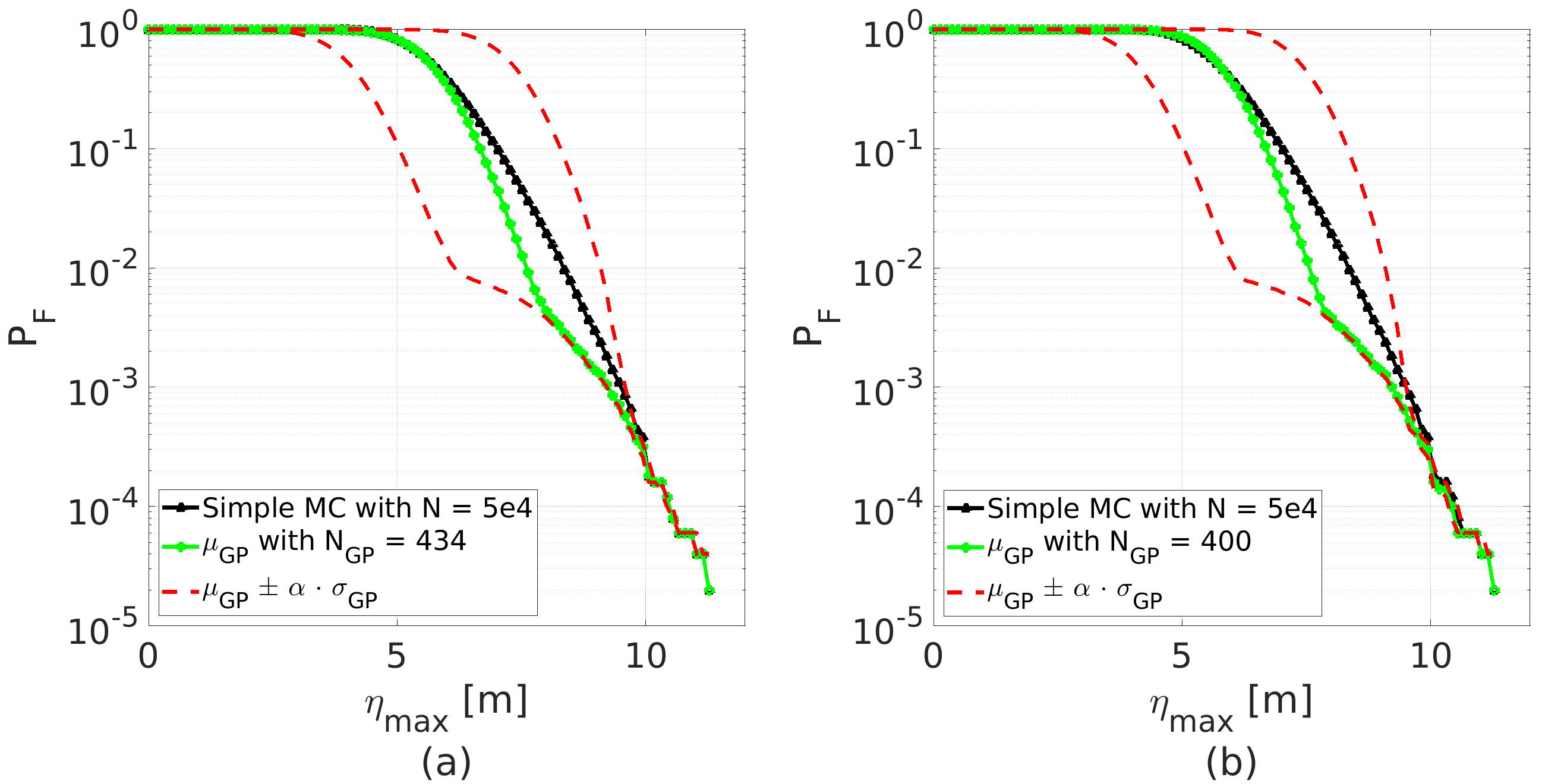}
    \caption{The final estimation of the short-term exceedance probability $\hat{P}_F$ for the KdV22 model using the Gaussian surrogate model with (a) the quadratic trend, and (b) the linear trend.}
    \label{kdv22:pfu}
\end{figure}

Figure \ref{kdv22:pfu} shows the final results after the learning process based on the $U$-design improvement (i.e., \textbf{Algorithm \ref{seq}}) has terminated after reaching the threshold $\xi=2$. For the pure quadratic trend, we have $\varepsilon^{\rm pure}_{9.45} = 1.66$ at $10^{-3}$, while $\varepsilon^{\rm lin}_{9.45} = 1$. The choice of the error threshold value of 2 seems reasonable in light of the typical levels of the exceedance probabilities (10$^{-4}$ or 10$^{-3}$). The pure quadratic trend requires more evaluations than the linear trend. However, the performance of the pure quadratic trend is better, with the mean squared error (MSE) less than $4\cdot10^{-4}$, while for the linear trend the MSE is less than $10^{-3}$. By visually inspecting Fig. \ref{kdv22:pfu} and Table \ref{kdv22:table}, it is observed that the model with the pure quadratic trend performs better. The total number of evaluations generated by the KdV22 for the pure quadratic trend is $N^{\rm pure}_{\rm GP}=434$ and for the linear trend $N^{\rm lin}_{\rm GP}=400$, which is around $0.87\%$ of the total number of simple Monte Carlo evaluations. Mostly, the design samples are closer to the events with higher peaks, see Fig. \ref{kdv22:data}. The simple Monte Carlo estimations are accurately recreated for extreme events with significant differences for $P_F > 10^{-3}$.

\begin{table}
\normalsize
\begin{adjustbox}{width=\columnwidth,center}
\begin{tabular}{c c c c c c} 
\hline\hline 
Method & $\hat{P}_F = 10^{-3}$ & $\hat{P}^+_F = 10^{-3}$ & $\hat{P}_F = 10^{-4}$  & $\hat{P}^+_F = 10^{-4}$ \\ [0.5ex] 
\hline 
Simple MC & 9.48 & - & 10.48 & - \\ 
Pure Quadratic & 9.38 & 9.67 & 10.48 & 10.54 \\ 
Linear & 9.22 & 9.55 & 10.42 & 10.54\\ [1ex]
\hline 
\end{tabular}
\end{adjustbox}
\caption{Estimation of the maximum wave crest $\eta_{\rm max}$ [m] using Gaussian process regression with the $U$-function for the KdV22 model and different exceedance levels.} 
\label{kdv22:table} 
\end{table}

\subsection{Application to fully nonlinear wave propagation over a slope}
We next apply the classification approach with active learning to compute the short-term exceedance probability for wave propagation over a slope with a fully nonlinear model, OceanWave3D \cite{allan}. OceanWave3D is a finite difference potential flow solver based on the Laplace equation with kinematic and dynamic free surface boundary conditions. An ad hoc wave breaking filter is included within the model. A complete derivation of the equations can be found in \cite{mc4}.

\begin{figure}[ht]
    \centering
    \includegraphics[scale=0.3]{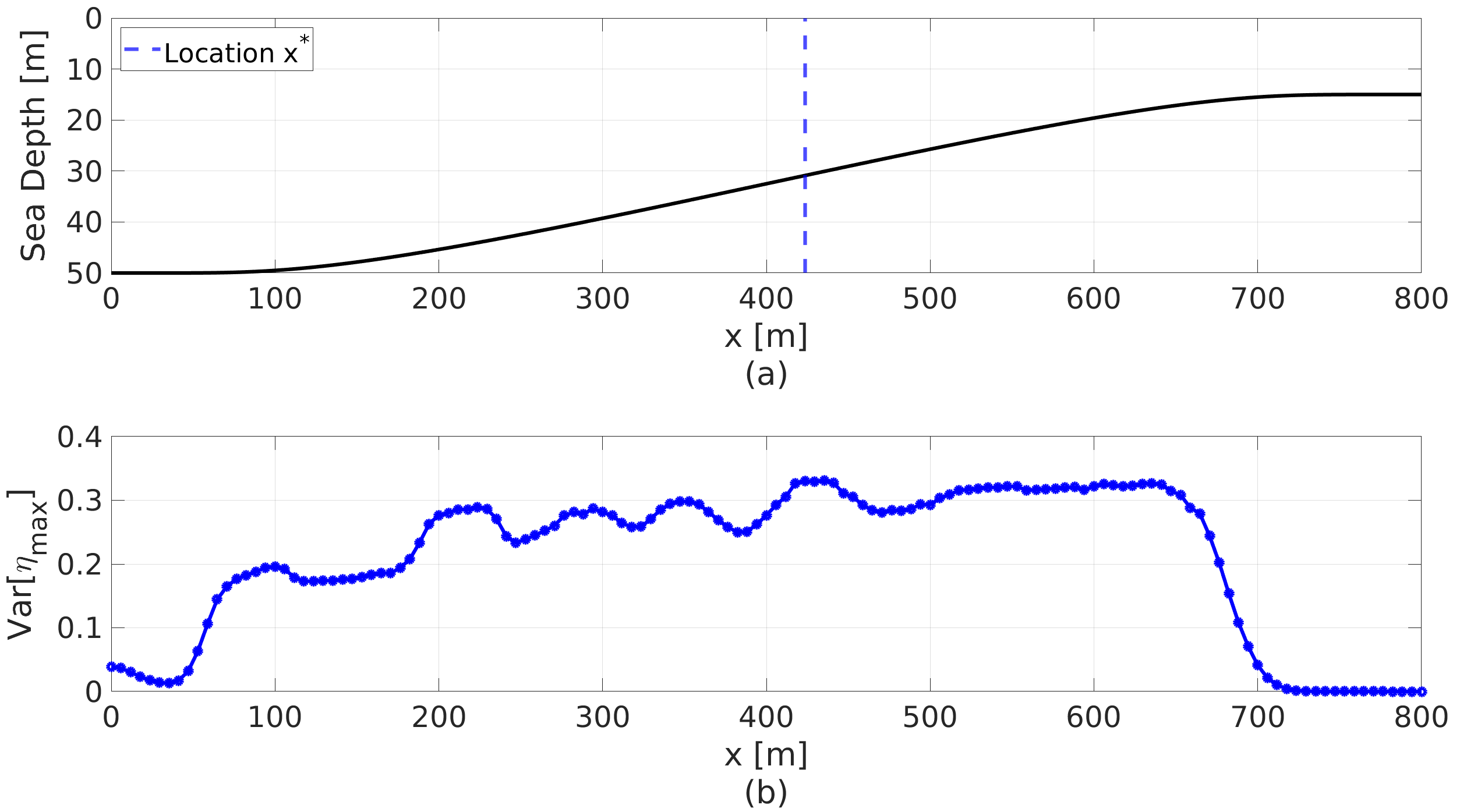}
    \caption{(a) Illustration of the sloping seabed with the reference location as the blue line. (b) The expectation of the squared deviation of $\eta_{\rm max}$ from its mean i.e. the variance for each spatial position.}
    \label{oc3d:bed}
\end{figure}

We establish our numerical experiment on the benchmark example of OceanWave3D defined for random waves over a slope \cite{bo}. MATLAB codes for this numerical experiment can be found at \url{https://github.com/ksehic/OCW3D-F90-UQProbe} in the folder ${\rm UQ-Packages}$. The sea state parameters for the JONSWAP spectrum are $H_S=4$m and $T_P=9$s. The computation is additionally simplified as it includes $137$ cells with a time duration of 256 seconds and 800 meters of the spatial domain. The initial sea depth is $50$m and gradually decreases to the shallow region of 15m. The reference location is at the sea depth of approximately $30$m, see Fig. \ref{oc3d:bed}a. Figure \ref{oc3d:bed}b shows the variance estimations for different spatial locations using the initial runs of OceanWave3D. We choose the reference location for the short-term exceedance probability $\hat{P}_F$ with the highest estimated variance. It is interesting to notice that the most uncertain location for this example and data is at the sea depth of $h\approx 30m$, which is a depth reduction of $40\%$.

\begin{figure}
    \centering
    \includegraphics[scale=0.24]{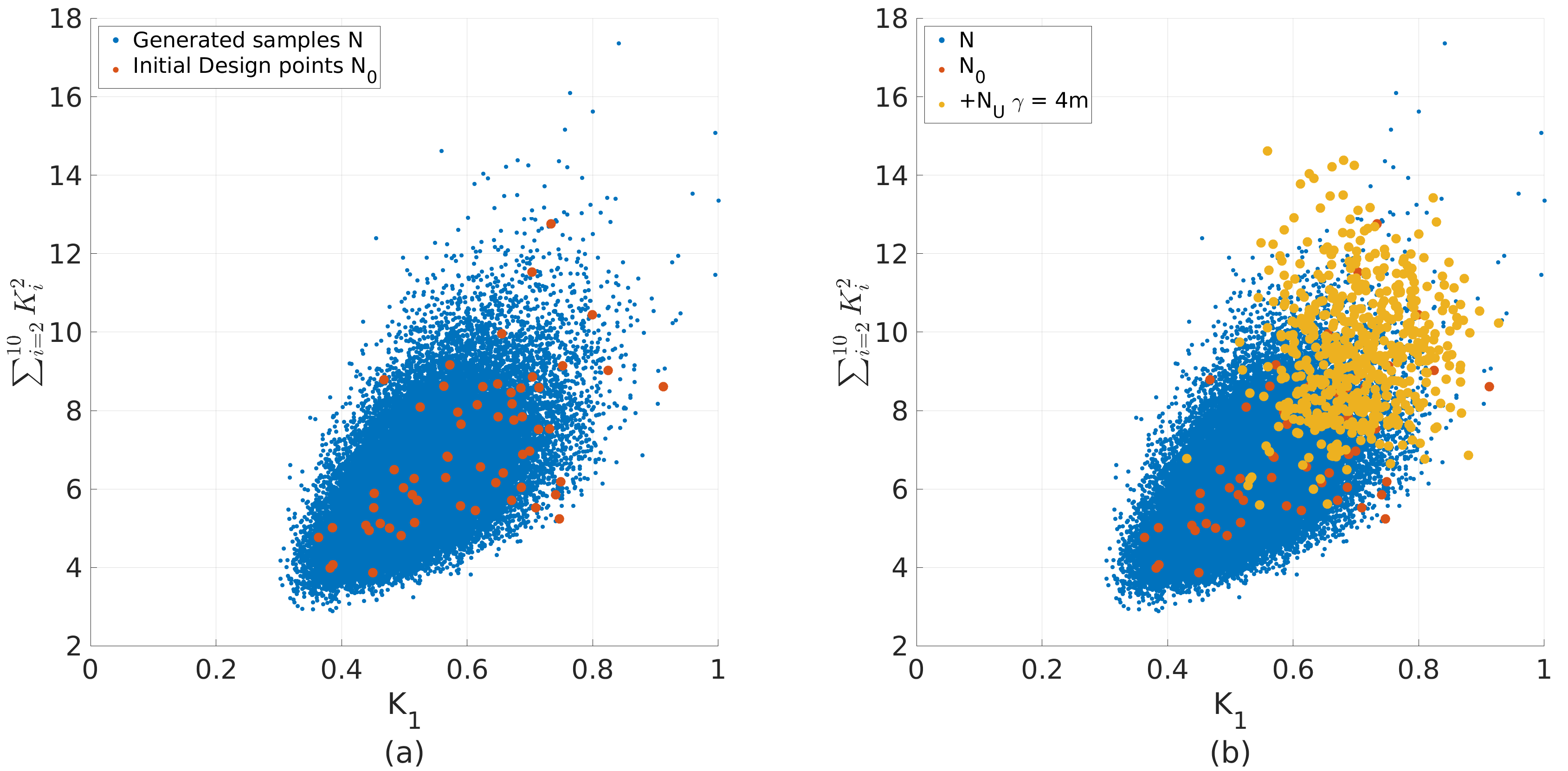}
    \caption{The wave generation classification data generated based on $H_S=4$m and $T_P=9$s with (a) initial training set $N_0$, and (b) additional $U$-samples $N_{\rm U}$ for $\gamma  = 4$m.}
    \label{oc3d:data}
\end{figure}

\begin{figure}
    \centering
    \includegraphics[scale=0.24]{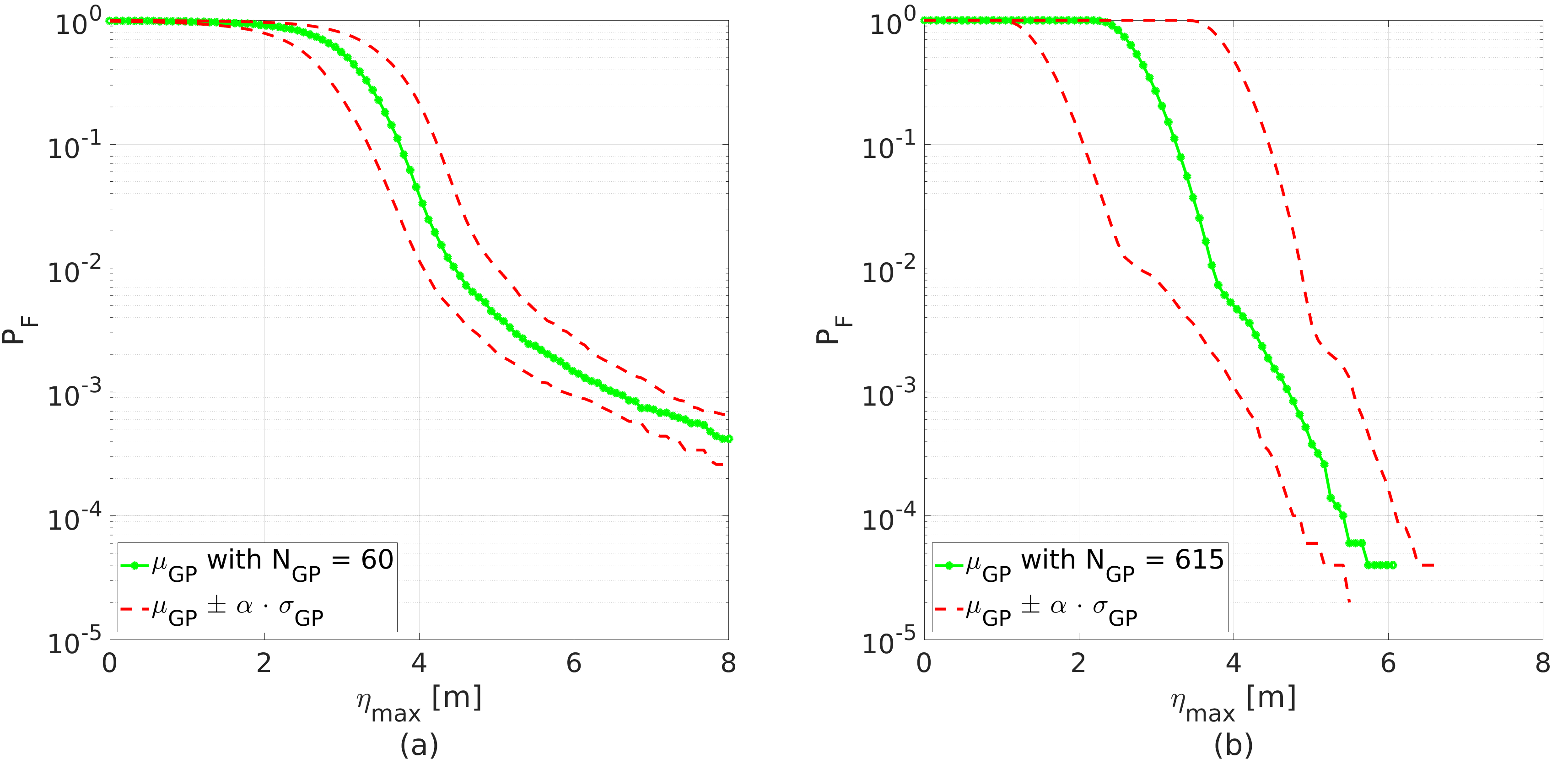}
    \caption{The estimation of the short-term exceedance probability $\hat{P}_F$ using the trained Gaussian process of OceanWave3D based on (a) $N_{\rm GP}=60$ initial design points, and (b) $N_{\rm GP} = 615$ with the additional $U$-design points for $\gamma=4$m.}
    \label{oc3d:pf0}
\end{figure}

\begin{figure}
    \centering
    \includegraphics[scale=0.24]{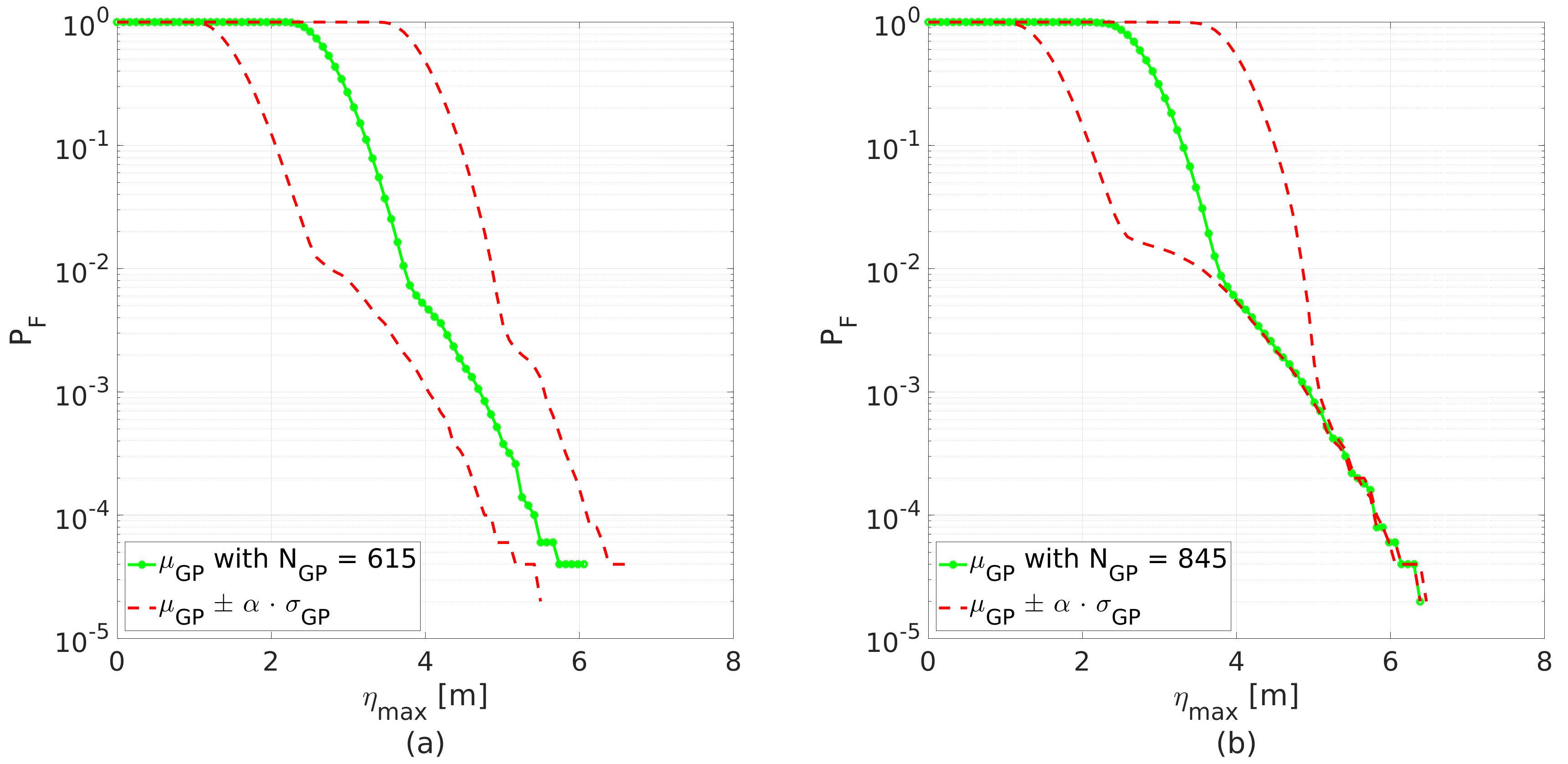}
    \caption{The estimation of the short-term exceedance probability $\hat{P}_F$ using the trained Gaussian process of OceanWave3D based on (a) $N_{\rm GP}=615$ training wave input samples for $\gamma=4$m, and (b) $N_{\rm GP} = 845$ with the additional $U$-design points after we change the failure threshold to $\gamma=4.6$m.}
    \label{oc3d:pf}
\end{figure}

The procedure is the same as previously described for the KdV22 shallow-water model. First, we generate and classify $N=5\cdot10^4$ wave input samples with the predefined sea state ($H_S = 4$m and $T_P=9$s) using $K_{1-10}$, see Fig. \ref{oc3d:data}. To define the training set, we randomly select initial $N_{\rm GP}=60$ wave input samples for which we evaluate OceanWave3D, see Fig. \ref{oc3d:data}a. Using this training set, we define the Gaussian process with the anisotropic squared exponential kernel function based on the pure quadratic regression. Next, we estimate the short-term exceedance probability as the sample mean with the indicator function using $N$ generated samples, see Fig. \ref{oc3d:pf0}a. Initially, we need to select a failure threshold for a numerical model and active learning. Thus, we select the failure threshold $\gamma = 4$m as $H_S$. The initial error measure Eq.~\eqref{er} attains the value $\varepsilon_4 = 5.4$, see Table \ref{oc:table}. This error measure is in this case too optimistic, in that the confidence interval is significantly narrowed with the error measure close to the predefined stopping criterion of $\xi = 2$. It is a clear example that one should consider the confidence interval of a Gaussian process with skepticism. We can see in Fig. \ref{oc3d:pf0}a that the tail of the distribution $P_F\geq10^{-2}$ is extremely heavy. The prediction is similar to the initial Gaussian prediction of the KdV22 model, see Fig. \ref{kdv22:pf}, which diverges significantly from its reference solution. If the prediction is true, additional carefully selected new wave input samples can only improve the stability of the Gaussian process. However, when the Gaussian process starts to learn and explore the probability space actively, the error measure $\varepsilon_4$ starts to converge to a value that is higher than the initial value. During the learning process (i.e., \textbf{Algorithm \ref{seq}}), in each iteration we add five new wave input samples with lowest $U$-values. Numerical evaluations can be run in parallel once optimal wave input samples are selected. After $615$ evaluations, we stop the algorithm and find the lowest error measure $\varepsilon_{4} = 95.7$ since the beginning of the learning process, see Fig. \ref{oc3d:pf}b. As expected, the wave input samples have mostly populated the area far from the origin, see Fig. \ref{oc3d:data}.

\begin{table}
\begin{adjustbox}{width=\columnwidth,center}
\begin{tabular}{c c c c c c} 
\hline\hline 
Training set & $\gamma = 3.45$m & $\gamma = 3.75$m & $\gamma = 4$m  & $\gamma = 4.6$m \\ [0.5ex] 
\hline 
$N_{\rm GP} = 60$ & 2 & 3.2 & 5.4 & 2.9\\ 
$N_{\rm GP} = 615$ & 23.9 & 90 & 95.7 & 32.4 \\ 
$N_{\rm GP} = 845$ & 30 & 1.1 & 0.1 & 0\\ [1ex]
\hline 
\end{tabular}
\end{adjustbox}
\caption{The convergence rate based on the error measure $\varepsilon_{\gamma}$, Eq.~\eqref{er}, at different failure levels $\gamma$ and training sets for OceanWave3D.} 
\label{oc:table} 
\end{table}

Visual inspection of Fig. \ref{oc3d:pf}a reveals that the Gaussian process reproduces the tail of the distribution. The Gaussian prediction of the short-term exceedance probability resembles typical offshore exceedance probabilities and is similar to what we have found with the KdV22 model. Still, the error measures are significant with high uncertainties in predictions, see Table \ref{oc:table2} for $N_{\rm GP} = 645$. We can see that the failure threshold for $10^{-3}$, which is our goal, will likely be higher than $4$m. Thus, we change the predefined failure threshold to $4.6$m as this threshold may correspond to the exceedance order of $10^{-3}$ eventually.

\begin{figure}
    \centering
    \includegraphics[scale=0.24]{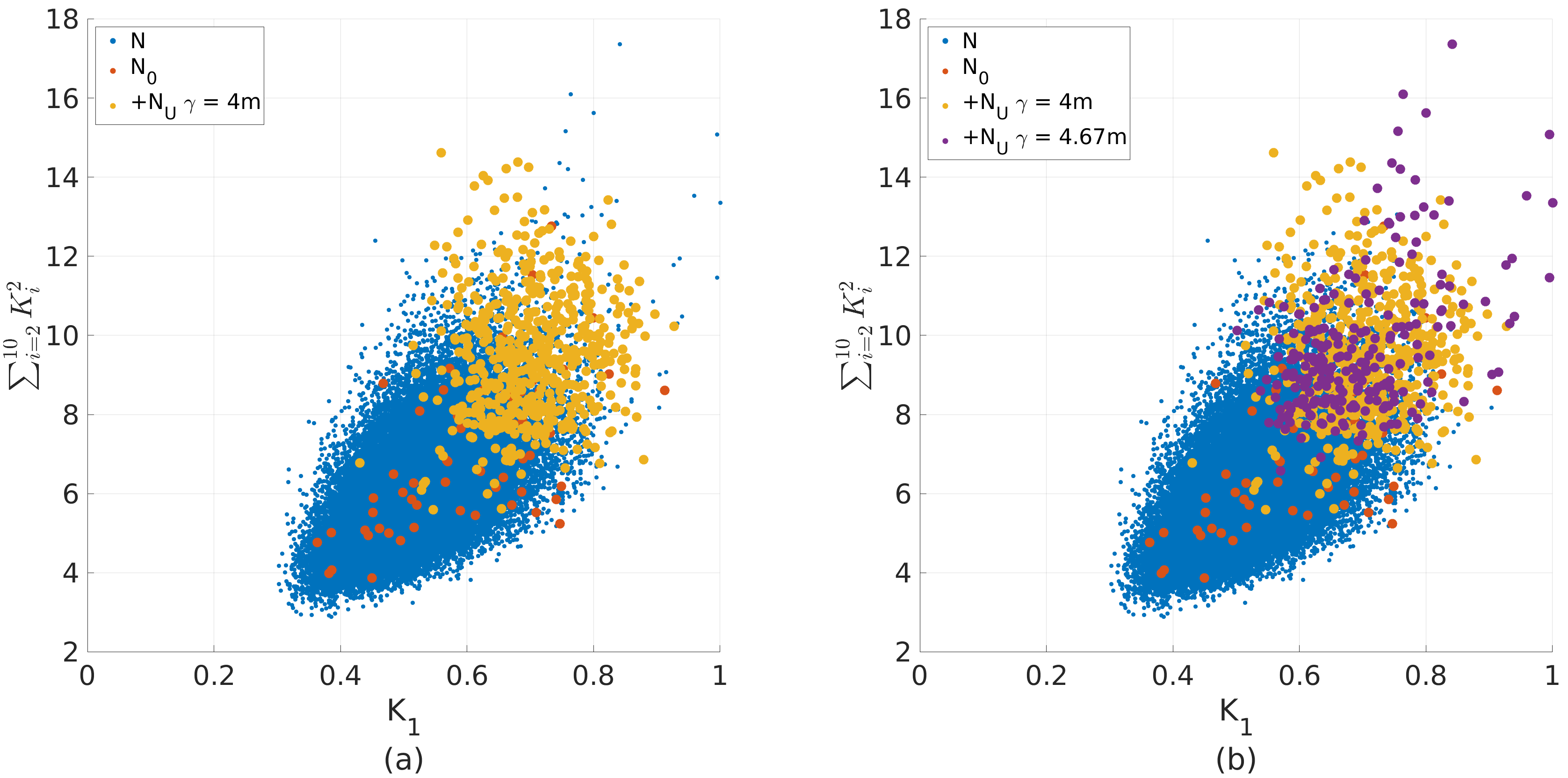}
    \caption{The samples $N_{\rm U}$ selected by the $U$-function with the failure threshold (a) $\gamma  = 4$m, and (b) $\gamma  = 4.67$m.}
    \label{oc3d:data1}
\end{figure}

We therefore iteratively include additional $230$ evaluations based on the $U$-function and the new failure threshold for the learning process, see Fig. \ref{oc3d:data1}b. The Gaussian process continues to learn and improve the predictions until the error measure drops below the stopping criterion of $\xi=2$, see Table \ref{oc:table2} for $N_{\rm GP} = 845$. The trained Gaussian process eventually estimates the short-term exceedance probability with the significantly narrowed confidence interval for the tail, see Fig. \ref{oc3d:pf}. The tail of the distribution is, to a certain extent, what we would expect for this case. With the new failure threshold $\gamma = 4.6$, the Gaussian process explores further in the tail, see Fig. \ref{oc3d:data1}b. 

In Table \ref{oc:table2}, we present the results for the maximum crest elevation $\eta_{\rm max}$ at the depth of approximately $30$m, for the exceedance probability levels $10^{-3}$ and $10^{-4}$. The maximum crest elevation for the exceedance level of $10^{-3}$ is $4.93$m. This is an increase of $23.2\%$ with respect to the chosen significant wave height $H_S=4$m. The upper bound for the same level is recorded as $5.05$m.

\begin{table}
\begin{adjustbox}{width=\columnwidth,center}
\begin{tabular}{c c c c c c} 
\hline\hline 
Training set & $\hat{P}_F = 10^{-3}$ & $\hat{P}^+_F = 10^{-3}$ & $\hat{P}_F = 10^{-4}$  & $\hat{P}^+_F = 10^{-4}$ \\ [0.5ex] 
\hline 
$N_{\rm GP} = 60$ & 6.54 & 7.15& 10.06 & 10.9 \\ 
$N_{\rm GP} = 615$ & 4.69 & 5.52& 5.41 & 6.1 \\ 
$N_{\rm GP} = 845$ & 4.93 & 5.05 & 5.8 & 5.81\\ [1ex]
\hline 
\end{tabular}
\end{adjustbox}
\caption{Estimation of the maximum wave crest $\eta_{\rm max}$ [m] using Gaussian process regression with the $U$-function for a fully nonlinear wave model, OceanWave3D.} 
\label{oc:table2} 
\end{table}

\begin{figure}
    \centering
    \includegraphics[scale=0.34]{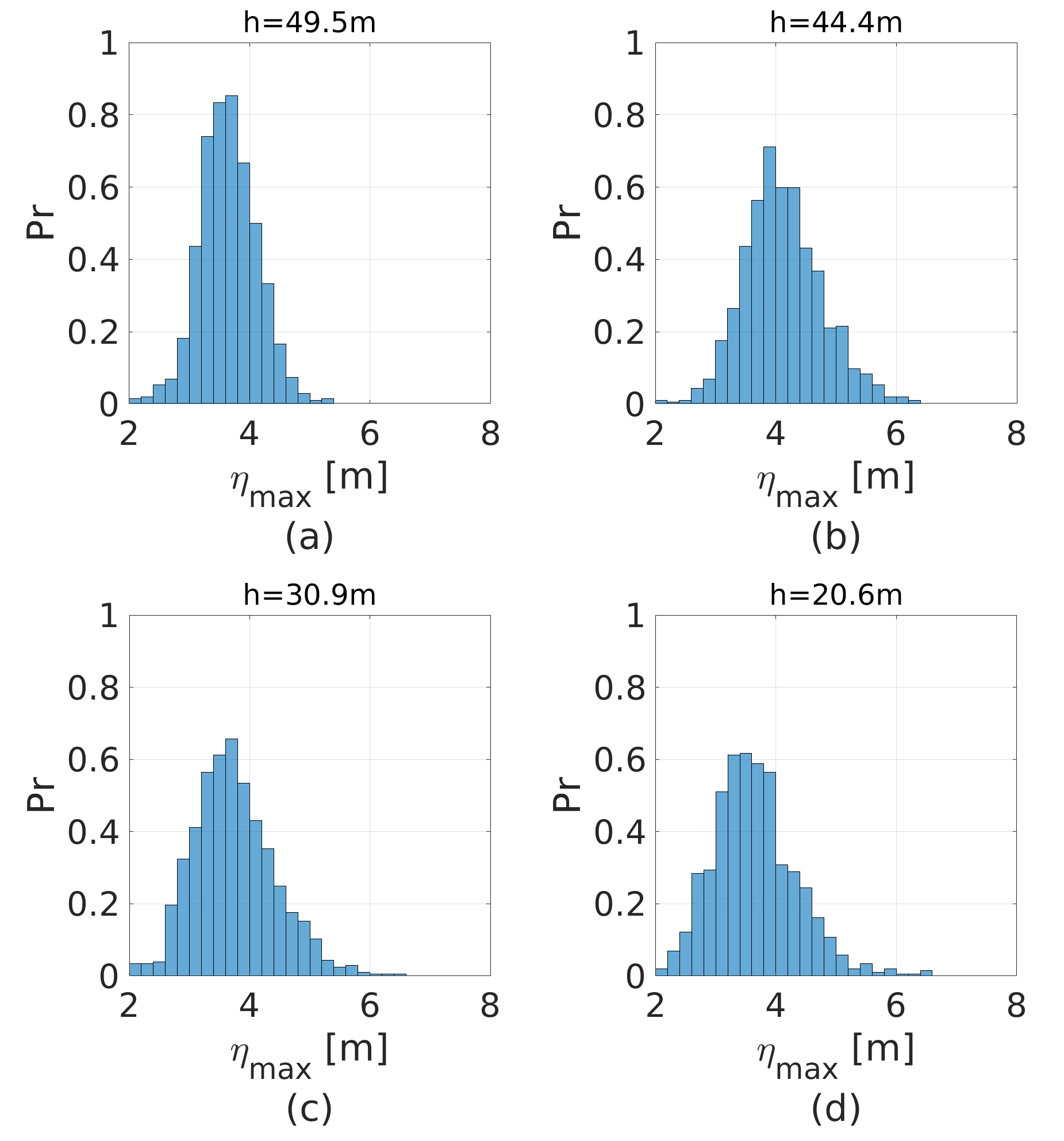}
    \caption{The histograms for different depths using $N_{\rm GP} = 845$ evaluations of OceanWave3D.}
    \label{oc3d:hist}
\end{figure}

Using $N_{\rm GP}=845$ training evaluations, we can additionally examine the effect of the slope on the quantity of interest and classification parameters. In Fig. \ref{oc3d:hist}, we plot the normalized histograms at four different depths. It is worth noticing how the slope, i.e., decreasing depth, causes the initially normal distribution at the depth of $50$m to change smoothly to a heavy-tailed distribution.

\begin{figure}
    \centering
    \includegraphics[scale=0.34]{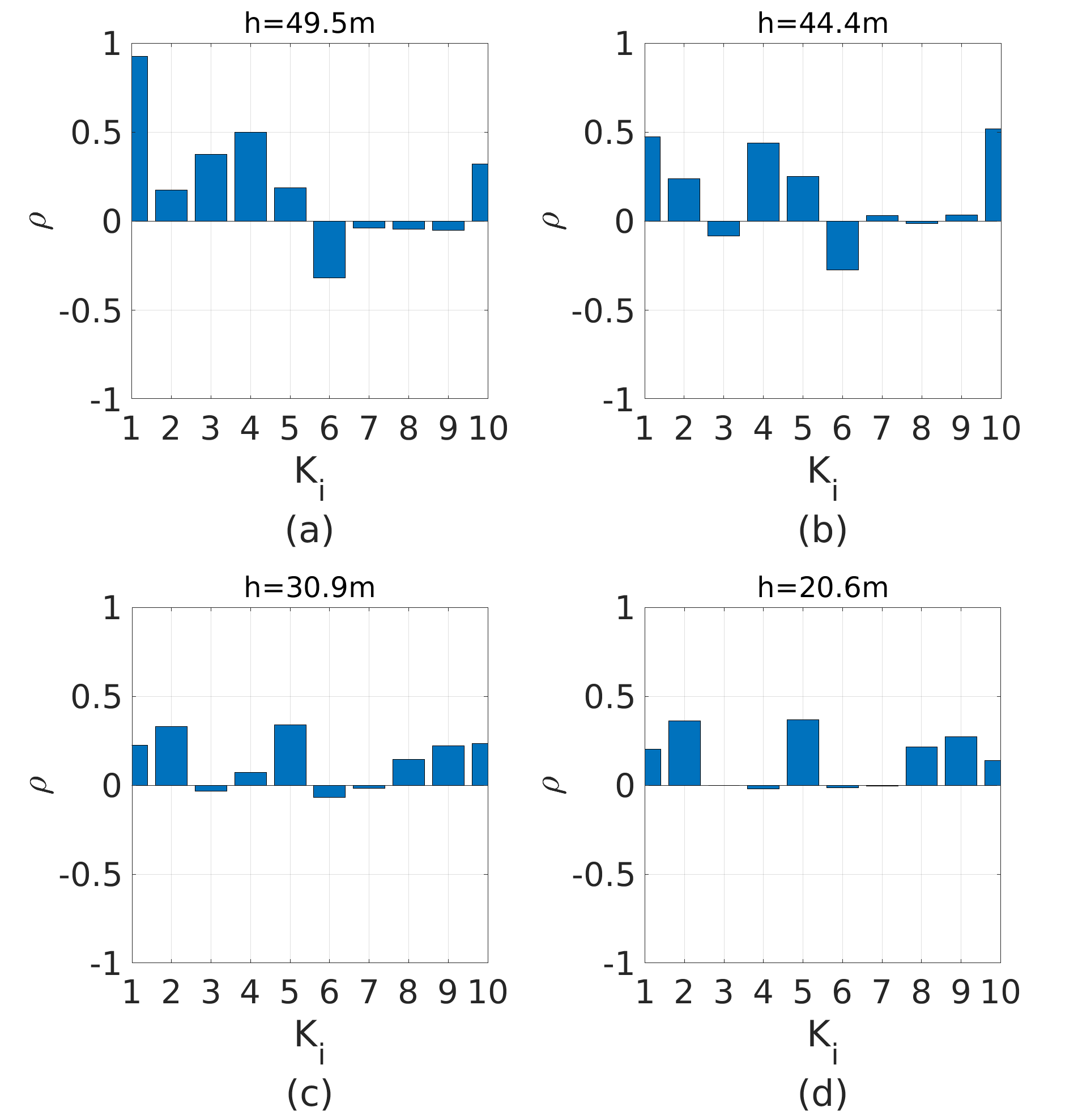}
    \caption{Correlation analysis for different depths using $N_{\rm GP} = 845$ evaluations of OceanWave3D.}
    \label{oc3d:corr}
\end{figure}

Figure \ref{oc3d:corr} describes the global sensitivity for the classification parameters $K_{1-10}$ concerning the quantity of interest $\eta_{\rm max}$ at four different depths using the Pearson correlation measure. While the seabed is flat, the classification parameter $K_1$ related to the maximum wave height at $x=0$, the kurtosis $K_4$, and approximated entropy $K_6$ are the most significant parameters. As the depth decreases, the parameters related to the variance $K_2$, such as the root mean square, the time-series percentiles for $75\%$ and $90\%$, and the variance $K_2$ itself influence the maximum crest elevation more than the rest of the parameters. It is interesting to notice how the skewness and kurtosis lose their influence, while the wave input percentiles for extreme events gain significance with decreasing depth. The approximate entropy $K_6$, which has a significant negative correlation with the variance, becomes negligible as the sea depth decreases. This parameter generally quantifies the regularity and unpredictability of time-series variation. 

\section{Conclusion and further work}\label{end}
When the high-dimensional input vector of Fourier coefficients is used to generate waves (i.e., time series), standard dimensionality reduction approaches cannot perform sufficiently to allow a quantification of extreme events with Gaussian process regression efficiently. Therefore, we propose to use time-series classification parameters (e.g., the variance or time-series percentages) as design parameters to determine a sufficient low-dimensional representation for a sequential design (i.e., active learning). Initially, we generated different independent realizations of surface elevation required for the simple Monte Carlo estimation of the short-term exceedance probability. By using the generated data, we defined the classification parameters and selected a few evaluations to train a Gaussian process. The initial Gaussian process estimated the short-term exceedance probability for the maximum crest elevation at the reference point, which deviated significantly from the reference solution. Thereafter, the surrogate model was improved actively using the $U$-function until the confidence interval of predictions dropped below some prescribed stopping criterion for the predefined failure threshold. The learning criterion based on the $U$-function emphasizes the predefined failure threshold and uncertain predictions. We demonstrate the applicability through two offshore problems, one of them involving a weakly nonlinear Korteweg-de Vries model to reproduce simple shallow-water wave conditions and other the wave propagation over a slope for a fully nonlinear model, OceanWave3D. Our proposal employs less than $1.7\%$ of the required Monte Carlo evaluations. As we based our approach on the generated data, the learning process is discrete and not optimal. The total number of evaluations can indeed be additionally reduced by continuous optimization. We plan to examine how to optimally prepare the data for learning and establish classification parameters for extreme events. Using the Pearson correlation measure, we can build and explore which classification parameters we should choose to quantify extreme events accurately and efficiently. In the present study, we use the stationary description of the wave input (i.e., a time-series). Therefore, we shall next focus on moving statistics and their influence on extreme events.

\section*{Acknowledgements}
This research was funded by the DeRisk project of Innovation Fund Denmark, grant number 4106-00038B.

\bibliographystyle{unsrt}  
\bibliography{references}  

\begin{thebibliography}{10}

\bibitem{bigoni}
D.~Bigoni, A.~P. Engsig-Karup, and C.~Eskilsson.
\newblock Efficient uncertainty quantification of a fully nonlinear and
  dispersive water wave model with random inputs.
\newblock {\em Journal of Engineering Mathematics}, 101(1):87--113, 2016.

\bibitem{samocita}
H.~Bredmose, M.~Dixen, A.~Ghadirian, T.~J. Larsen, S.~Schløer, S.~J. Andersen,
  S.~Wang, H.~B. Bingham, O.~Lindberg, E.D. Christensen, M.~H. Vested,
  S.~Carstensen, A.~P. Engsig-Karup, O.~S. Petersen, H.~F. Hansen, J.~S.
  Mariegaard, P.~H. Taylor, T.~A.~A. Adcock, C.~Obhrai, O.~T. Gudmestad, N.~J.
  Tarp-Johansen, C.~P. Meyer, J.~R. Krokstad, L.~Suja-Thauvin, and T.~D.
  Hanson.
\newblock De{R}isk — accurate prediction of {ULS} wave loads. {O}utlook and
  {F}irst {R}esults.
\newblock {\em Energy Procedia}, 90:379--387, 2016.

\bibitem{samocita2}
A.~Ghadirian and H.~Bredmose.
\newblock Pressure impulse theory for a slamming wave on a vertical circular
  cylinder.
\newblock {\em Journal of Fluid Mechanics}, 867:R1, 2019.

\bibitem{turk}
B.~Yildirim and G.~E. Karniadakis.
\newblock Stochastic simulations of ocean waves: An uncertainty quantification
  study.
\newblock {\em Ocean Modelling}, 86:15--35, 2015.

\bibitem{form}
R.~Rackwitz.
\newblock Reliability analysis - a review and some perspectives.
\newblock {\em Structural Safety}, 23:365--395, 2001.

\bibitem{mcbook}
A.~B. Owen.
\newblock {\em Monte Carlo theory, methods and examples}.
\newblock Open Access, 2013.

\bibitem{xiuli}
J.~Li and D.~Xiu.
\newblock Evaluation of failure probability via surrogate models.
\newblock {\em Journal of Computational Physics}, 229(23):8966--8980, 2010.

\bibitem{bruno}
R.~Sch\"obi, B.~Sudret, and S.~Marelli.
\newblock Rare event estimation using polynomial-chaos kriging.
\newblock {\em Journal of Risk and Uncertainty in Engineering Systems, Part A:
  Civil Engineering, American Society of Civil Engineers (ASCE)}, 3(2), 2016.

\bibitem{paul1}
P.~G. Constantine.
\newblock {\em Active Subspaces: Emerging Ideas for Dimension Reduction in
  Parameter Studies}.
\newblock Society for Industrial and Applied Mathematics, 2015.

\bibitem{evt}
M.~Nicodemi.
\newblock {\em Extreme value statistics}.
\newblock Springer New York, 2012.

\bibitem{ld2}
G.~Dematteis, T.~Grafke, and E.~Vanden-Eijnden.
\newblock Rogue waves and large deviations in deep sea.
\newblock {\em Proceedings of the National Academy of Sciences},
  115(5):855--860, 2018.

\bibitem{ld}
S.~R.~S. Varadhan.
\newblock {\em Large Deviations and Applications}.
\newblock Society for Industrial and Applied Mathematics, 1984.

\bibitem{fpe}
H.~Risken.
\newblock {\em The Fokker-Planck Equation Methods of Solution and
  Applications}.
\newblock Springer New York, 2 edition, 1989.

\bibitem{mo}
M.~A. Mohamad and T.~P. Sapsis.
\newblock Sequential sampling strategy for extreme event statistics in
  nonlinear dynamical systems.
\newblock {\em Proceedings of the National Academy of Sciences},
  115(44):11138--11143, 2018.

\bibitem{sapsis}
W.~Cousins and T.~P. Sapsis.
\newblock Reduced-order precursors of rare events in unidirectional nonlinear
  water waves.
\newblock {\em Journal of Fluid Mechanics}, 790:368--388, 2016.

\bibitem{asa}
K.~\v{S}ehi\'{c}, H.~Bredmose, J.~S{\o}rensen, and M.~Karamehmedovi\'c.
\newblock Active-subspace analysis of exceedance probability for shallow-water
  waves.
\newblock {\em TBD}, TBD:TBD, TBD.

\bibitem{wavepack2}
P.~Boccotti.
\newblock Some new results on statistical properties of wind waves.
\newblock {\em Applied Ocean Research}, 5(3):134--140, 1983.

\bibitem{wavepack1}
G.~Lindgren.
\newblock Some properties of a normal process near a local maximum.
\newblock {\em The Annals of Mathematical Statistics}, 41(6):1870--1883, 1970.

\bibitem{wavepack3}
P.~S. Tromans, A.~R. Anaturk, and A.~Hagemeijer.
\newblock A new model for the kinematics of large ocean waves-application as a
  design wave.
\newblock {\em Proceeding of The First International Offshore and Polar
  Engineering Conference}, 3:64--71, 1991.

\bibitem{pcabook}
I.T. Jolliffe.
\newblock {\em Principal Component Analysis}.
\newblock Springer, second edition, 2002.

\bibitem{pls1}
M.A. Bouhlel, N.~Bartoli, A.~Otsmane, and J.~Joseph~Morlier.
\newblock Improving kriging surrogates of high-dimensional design models by
  partial least squares dimension reduction.
\newblock {\em Structural and Multidisciplinary Optimization}, 53:935--952,
  2016.

\bibitem{pls4}
I.~Papaioannou, M.~Ehre, and D.~Straub.
\newblock {PLS}-based adaptation for efficient {PCE} representation in high
  dimensions.
\newblock {\em Journal of Computational Physics}, 387:186--204, 2019.

\bibitem{vae_var}
F.~J. Gonzalez and M.~Balajewicz.
\newblock Deep convolutional recurrent autoencoders for learning
  low-dimensional feature dynamics of fluid systems.
\newblock {\em arXiv:1808.01346}, 2018.

\bibitem{allan}
A.~P. Engsig-Karup, H.~B. Bingham, and O.~Lindberg.
\newblock An efficient flexible-order model for {3D} nonlinear water waves.
\newblock {\em Journal of Computational Physics}, 228:2100--2118, 2009.

\bibitem{rasmussen}
C.~E. Rasmussen and C.~K. I.Williams.
\newblock {\em Gaussian Processes for Machine Learning}.
\newblock MIT Press, 2006.

\bibitem{gramacy}
R.~B. Gramacy and D.~W. Apley.
\newblock Local gaussian process approximation for large computer experiments.
\newblock {\em Journal of Computational and Graphical Statistics}, 24:561--578,
  2015.

\bibitem{naess}
A.~Naess and T.~Moan.
\newblock {\em Stochastic Dynamics of Marine Structures}.
\newblock Cambridge University Press, 2012.

\bibitem{low}
B.~Esmael, A.~Arnaout, R.~K. Fruhwirth, and G.~Thonhauser.
\newblock A statistical feature-based approach for operations recognition in
  drilling time series.
\newblock {\em International Journal of Computer Information Systems and
  Industrial Management Applications}, 5(2150--7988):454--461, 2013.

\bibitem{bo}
B.~T. Paulsen.
\newblock {OceanWave3D}.
\newblock \url{https://github.com/boTerpPaulsen/OceanWave3D-Fortran90}, Dec
  2019.

\bibitem{henrik}
H.~Bredmose.
\newblock Evolution equations for wave-wave interaction.
\newblock Master's thesis, Technical University of Denmark, Lyngby, Denmark,
  1999.

\bibitem{Conrad:2016}
P.~R. Conrad, Y.~M. Marzouk, N.~S. Pillai, and A.~Smith.
\newblock {A}ccelerating {A}symptotically {E}xact {MCMC} for {C}omputationally
  {I}ntensive {M}odels via {L}ocal {A}pproximations.
\newblock {\em Journal of the American Statistical Association},
  111:516:1591--1607, 2016.

\bibitem{mc4}
A.~P. Engsig-Karup, L.~S. Glimberg, A.~S. Nielsen, and O.~Lindberg.
\newblock Fast hydrodynamics on heterogenous many-core hardware.
\newblock In Rapha\"el Couturier, editor, {\em Designing Scientific
  Applications on GPUs}, chapter~11, pages 251--294. Taylor \& Francis, 2013.

\end{thebibliography}

\end{document}